\documentclass[prl,twocolumn,showpacs,preprintnumbers,amsmath,amssymb,superscriptaddress]{revtex4}

\usepackage{graphicx, color, graphpap}
\usepackage{amsmath}
\usepackage{amssymb}
\usepackage{amsthm}
\usepackage{pstricks}
\usepackage{float}
\usepackage{multirow}
\usepackage{hyperref}

\long\def\ca#1\cb{} 

\newcommand{\ket}[1]{|#1\rangle}               
\newcommand{\colo}{\,\hbox{:}\,}              
\newcommand{\bra}[1]{\langle #1|}              
\newcommand{\dya}[1]{\ket{#1}\bra{#1}}
\newcommand{\dyad}[2]{\ket{#1}\bra{#2}}        
\newcommand{\ip}[2]{\langle #1|#2\rangle}      

\newcommand{\EC}{\mathcal{E}}

\newcommand{\HC}{\mathcal{H}}

\newcommand{\Tr}{{\rm Tr}}
\renewcommand{\geq}{\geqslant}
\renewcommand{\leq}{\leqslant}

\newcommand{\ot}{\otimes}
\newcommand{\ad}{^\dagger}

\newcommand{\al}{\alpha }
\newcommand{\bt}{\beta }
\newcommand{\gm}{\gamma }

\newcommand{\dl}{\delta }

\newcommand{\lm}{\lambda }
\newcommand{\Lm}{\Lambda }

\newcommand{\sg}{\sigma }

\newcommand{\om}{\omega }
\newcommand{\Om}{\Omega }

\newtheoremstyle{example}{\topsep}{\topsep}%
{}
{}
{\bfseries}
{.}
{   }
{\thmname{#1}\thmnumber{ #2}}
\theoremstyle{example}
\newtheorem{example}{Example}

\newtheorem*{thm4}{Lemma S1}
\newtheorem{theorem}{Theorem}

\newtheorem{corollary}[theorem]{Corollary}

\begin{document}

\title{Relative entropy derivation of the uncertainty principle with quantum side information}
\author{Patrick J. Coles}
\affiliation{Department of Physics, Carnegie Mellon University, Pittsburgh, Pennsylvania 15213, USA}
\author{Li Yu}
\affiliation{Department of Physics, Carnegie Mellon University, Pittsburgh, Pennsylvania 15213, USA}
\author{Michael Zwolak}
\affiliation{Theoretical Division, Los Alamos National Laboratory, Los Alamos, NM 87545}

\begin{abstract}
We give a simple proof of the uncertainty principle with quantum side information, as in [Berta et al.\ Nature Physics 6, 659 (2010)], invoking the monotonicity of the relative entropy. Our proof shows that the entropic uncertainty principle can be viewed as a data-processing inequality, a special case of the notion that information cannot increase due to evolution in time. This leads to a systematic method for finding the minimum uncertainty states of various entropic uncertainty relations; interestingly such states are intimately connected with the reversibility of time evolution.
\end{abstract}

\maketitle

Classical information theory, pioneered by Shannon \cite{Shannon}, addresses the question of how information storage, processing, and transmission tasks can be performed with macroscopic, \emph{decohered} resources. The more general question of what can be done with resources that may or may not be decohered is the subject of quantum information theory. All of Shannon's quantitative tools, such as entropy and mutual information, apply perfectly well in the quantum domain provided that one focuses on a \emph{single} type of information \cite{Gri07}, associated with a particular measurement on the quantum system of interest. What makes quantum information theory \emph{different} from its classical counterpart is precisely the existence of multiple types of information or properties of a quantum system, e.g.\ the $x$ and $z$ components of an electron's spin, and the notion that these properties can be \emph{incompatible} in the sense that one cannot simultaneously know both types of information. This purely quantum idea is captured quantitatively in the uncertainty principle.

Formulations of the uncertainty principle have become progressively stronger over the years. Variances \cite{Robertson} have been replaced by entropies \cite{EURreview1} as measures of uncertainty, and recent formulations allow the observer to possess background or ``side" information about the quantum observables, i.e.\ either classical \cite{Hall1} or quantum \cite{RenesBoileau,BertaEtAl} side information. These latter formulations for two bases, and their generalization to two POVMs \cite{TomRen2010,ColesEtAlv4} and to smooth entropies \cite{BertaEtAl,TomRen2010}, represent the strongest versions of the uncertainty principle for two observables to date. 

Thusfar, the uncertainty principle with quantum side information (UPQSI) in terms of Shannon entropies has only been proven as a corollary to a similar formulation in terms of smooth entropies \cite{BertaEtAl, TomRen2010}, so there is the question as to whether the machinery of smooth entropies is necessary to understand the UPQSI. While smooth entropies have operational meanings \cite{KonRenSch09} and show great promise for quantum cryptography \cite{TomRen2010}, one still yearns for the \emph{intuition} behind the UPQSI. In this article, we derive the UPQSI using the properties of the relative entropy, which plays a central role in quantum information theory \cite{VedralReview02} and is, thus, familiar to many researchers in this field. In particular, we find that the UPQSI is connected to the fact that the relative entropy, which roughly acts like a distance between two density operators, does not increase over time, a principle called the monotonicity of the relative entropy. This approach allows us to generalize the UPQSI to a state-dependent bound, which strengthens it when the measurement(s) are complementary to one's prior knowledge of the state.

This approach also leads us to a systematic method for answering the question: when is the uncertainty principle satisfied with \emph{equality}? Such states are called minimum uncertainty states (MUS). Squeezed states of the harmonic oscillator, which have application in high-sensitivity interferometry and gravity-wave detection \cite{BonSha84,Walls83}, are well-known MUS of the variance uncertainty relation \cite{Robertson}. Very little is known about the MUS of \emph{entropic} uncertainty relations, though see \cite{RenesBoileau,Renes2010}. Knowledge of such MUS may help in optimizing recently proposed applications of the UPQSI to entanglement witnessing and quantum cryptography \cite{BertaEtAl}. In this article we find necessary and sufficient conditions for a state to be a MUS, for several entropic uncertainty relations.

\textit{Conditional entropy}. The uncertainty or missing information about a POVM $P_a=\{P_{a,j}\}$ on system $a$ is given by Shannon's entropy of the associated probability distribution $\{p_j\}$: $H(P_a)= H(\{p_j\}) = -\sum_j p_j\log p_j$. Classical side information, e.g.\ given by a POVM $Q_b$ on system $b$, only reduces one's uncertainty about $P_a$: 
\begin{equation}
\label{eqn1}
 H(P_a|Q_b)=H(P_a)-H(P_a\colo Q_b)\leq H(P_a)
\end{equation}
where $H(P_a\colo Q_b)=H(P_a)+H(Q_b)-H(P_a,Q_b)$ is the mutual information. A quantum analog:
\begin{equation}
   H_{}(P_a|b):=H(P_a)-\chi(P_a,b),
\label{eqn2}
\end{equation}
comes from replacing $H(P_a\colo Q_b)$ in \eqref{eqn1} with the Holevo quantity $\chi(P_a,b)=S(\rho_b)-\sum_j p_j S(\rho_{b,j})$, where $\rho_b=\Tr_a(\rho_{ab})$, $\rho_{b,j}=\Tr_a(P_{a,j}\rho_{ab})/p_j$, $\rho_{ab}$ is the quantum state of $ab$, and $S(\rho)=-\Tr (\rho \log \rho)$ is von Neumann's entropy. By Holevo's bound $ H_{}(P_a|b)\leq H(P_a|Q_b)$, and by analogy to $H(P_a\colo Q_b)$ we say that $\chi(P_a,b)$ measures the ``quantum side information" \cite{DevWin03} about $P_a$ located in system $b$. Also, $ H_{}(P_a|b)\geq 0$ and equals zero iff $b$ perfectly contains the $P_a$ information \cite{ColesEtAlv4}. Henceforth we drop the $a$ subscript from $P_a$. We note that another quantum analog of \eqref{eqn1} is $S(a|b)=S(\rho_{ab})-S(\rho_b)$, which can be negative for entangled $\rho_{ab}$.

\begin{figure}[t]
\begin{center}
\includegraphics[scale=0.36]{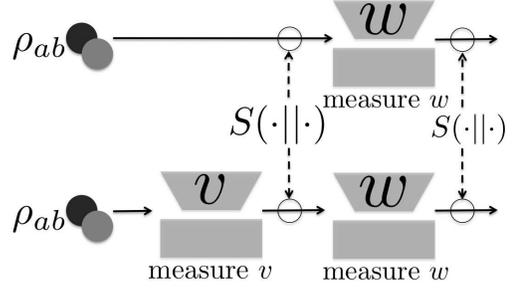}
\caption{%
  The UPQSI states that the relative entropy after the $w$ measurement is never larger than that just before it.\label{fgr1}}
\end{center}
\end{figure}

\textit{Uncertainty relation for bases}. The UPQSI strongly constrains the possible correlations in a tripartite state $\rho_{abc}$, stating that if $b$ knows something about an observable of $a$, then $c$ cannot know too much about about a complementary observable of $a$. The proof of the UPQSI is simplest for two bases $v=\{\ket{v_j}\}$ and $w=\{\ket{w_k}\}$ of $\HC_a$ that are mutually unbiased bases (MUBs): $|\ip{v_j}{w_k}|^2=1/d$ for all $j,k$, where $d$ is the dimension of $\HC_a$. We wish to show that:
\begin{equation}
\label{eqn3}
 H_{}(v|c)+ H_{}(w|b)\geq \log d,
\end{equation} 
noting that the proof for pure $\rho_{abc}$ immediately implies the proof for mixed $\rho_{abc}$ by the concavity of conditional entropy (p.~520 of \cite{NieChu00}). We exploit the connection, proved in \cite{ColesEtAlv4}, between the conditional entropy and \textit{relative entropy} $S(\rho||\sg)= \Tr(\rho\log \rho)-\Tr(\rho\log \sg)$: let $\rho_{abc}$ be a pure state, then
\begin{equation}
\label{eqn4}
 H_{}(v|c)=S(\rho_{ab}||\sum_j [v_j]\rho_{ab}[v_j]),
\end{equation} 
where we use the notation $[\psi]:=\dya{\psi}$ for a rank-1 projector \cite{CQT} \footnote{We remark that \eqref{eqn4} is connected to the quantum discord \cite{OllZur01} since $S(\rho_{ab}||\sum_j [v_j]\rho_{ab}[v_j])$ measures the ``distance" to a zero-discord state; we investigate this in future work.}. It follows that
\begin{align}
\label{eqn5}& H_{}(v|c)=S(\rho_{ab}||\sum_j [v_{j}]\ot \Tr_a\{[v_{j}]\rho_{ab}\})\\
&\geq S(\sum_k [w_{k}]\ot \Tr_a\{[w_{k}]\rho_{ab}\}||\nonumber\\
\label{eqn6}&\sum_{j,k} |\ip{v_{j}}{w_{k}}|^2[w_{k}]\ot \Tr_a\{[v_j]\rho_{ab}\})\\
\label{eqn7}&= S(\sum_k [w_{k}]\ot \Tr_a\{[w_{k}]\rho_{ab}\}|| (I_a/d)\ot \rho_{b})\\
\label{eqn8}&= \log d+S(\sum_k [w_{k}]\ot \Tr_a\{[w_{k}]\rho_{ab}\} || I_a\ot \rho_{b})\\
\label{eqn9}&= \log d - H_{}(w|b).
\end{align}
Step~\eqref{eqn6} invoked $S(\rho||\sg)\geq S(\EC(\rho)||\EC(\sg))$ \cite{VedralReview02} with $\EC(\rho)=\sum_k [w_k]\rho [w_k]$, \eqref{eqn8} invoked $S(\rho|| \bt \sg)=S(\rho ||\sg)-\log \bt $ for some positive number $\bt$, and \eqref{eqn9} invoked Eq.~(11.58) of \cite{NieChu00}. A schematic diagram of this proof is shown in Fig.~\ref{fgr1}.

Now consider two arbitrary bases $v$ and $w$, with $r(v,w)=\max_{j,k} |\ip{v_{j}}{w_{k}}|^2$. We wish to prove the main inequality from \cite{BertaEtAl}:
\begin{equation}
\label{eqn10}
 H_{}(v|c)+ H_{}(w|b)\geq -\log r(v,w).
\end{equation} 
We follow the same strategy as for MUBs, from \eqref{eqn6}:
\begin{align}
H_{}(v|c)&\geq S(\sum_k [w_{k}]\ot \Tr_a\{[w_k]\rho_{ab}\} ||r(v,w)I_a\ot\rho_b)\nonumber \\
&=-  H_{}(w|b) - \log r(v,w). \nonumber
\end{align}
Here we used the fact \cite{OhPe93} that $S(\rho||\sg)\geq S(\rho ||\tau)$ if $\tau \geq \sg$; i.e.\ replacing each $|\ip{v_{j}}{w_{k}}|^2$ in \eqref{eqn6} with $r(v,w)$ makes the overall operator larger.

While it is clear that \eqref{eqn10} implies the well-known uncertainty relation of Maassen and Uffink \cite{MaassenUffink}: 
\begin{equation}
\label{eqn11}
 H_{}(v)+ H_{}(w)\geq -\log r(v,w),
\end{equation} 
one can \emph{directly} prove \eqref{eqn11} starting with $H(v)=S([\psi]||\sum_j [v_{j}][\psi] [v_{j}])$ for a pure state $\ket{\psi}\in\HC_a$, and proceeding with the same sort of steps shown above; a proof that is simpler than the original \cite{MaassenUffink}.

\textit{State-dependent bound}. More generally the UPQSI can be written for two POVMs $P$ and $Q$ \cite{TomRen2010,ColesEtAlv4}:
\begin{equation}
\label{eqn12}
 H_{}(P|b)+ H_{}(Q|c)\geq -\log r(P,Q).
\end{equation} 
where $r(P,Q)=\max_{j,k} \| \sqrt{Q_{k}} \sqrt{P_{j}}\|_\infty^2$ and $\| \cdot \|_\infty$ denotes the supremum norm (the maximum singular value). One can even formulate an UPQSI for a \emph{single} POVM \cite{ColesEtAlv4}:
\begin{equation}
\label{eqn13}
 H_{}(P|b)\geq -\log \max_{j} \| P_{j}\|_\infty.
\end{equation} 
Here we generalize \eqref{eqn12} and \eqref{eqn13}, replacing the right-hand-sides with (possibly) state-dependent bounds \footnote{See supplemental material for proofs of Theorems~\ref{thm1}--\ref{thm5} and elaboration on the MUS of \eqref{eqn23}.}.

\begin{theorem}
\label{thm1}
Let $P=\{P_{j}\}$ and $Q=\{Q_{k}\}$ be arbitrary POVMs, and let $\Pi$ be any projector on $a$ that projects onto a space that contains the support of $\rho_a$, then
\begin{equation}
\label{eqn14}
 H_{}(P|b)+ H_{}(Q|c)\geq -\log r(P,Q; \Pi),
\end{equation} 
where $r(P,Q; \Pi)=\max_{j,k} \| \sqrt{Q_{k}} \Pi \sqrt{P_{j}}\|_\infty^2$, and each $H(\cdot |\cdot)$ term is bounded by, e.g.\
\begin{equation}
\label{eqn15}
 H_{}(P|b)\geq -\log \max_{j} \|\Pi \sqrt{P_{j}}\|_\infty^2.
\end{equation} \openbox
\end{theorem}

Setting $\Pi=I_a$ reduces \eqref{eqn14} and \eqref{eqn15} to \eqref{eqn12} and \eqref{eqn13}. In many cases choosing a $\Pi$ with a lower rank than $I_a$ in \eqref{eqn14} leads to a stronger bound (examples below), though this is not a general rule. On the other hand, the strongest bound in \eqref{eqn15} always results from chosing $\Pi$ to have the smallest possible rank, i.e.\ the projector onto the support of $\rho_a$ (see \footnotemark[\value{footnote}]). We remark that all the results in \cite{ColesEtAlv4} hold if one replaces $r(P,Q)$ with $r(P,Q;\Pi)$. For example, if $P$ is any POVM on $a$ and $N$ is a rank-1 POVM on $a$, then
\begin{equation}
\label{eqn16}
 H_{}(P|b)+ H_{}(N|b)\geq -\log r(P,N;\Pi) +S(a|b),
\end{equation}
which is obtained from \eqref{eqn14} applied to pure $\rho_{abc}$ by adding $H(N|b)-H(N|c)=S(a|b)$ (see \cite{ColesEtAlv4}) to both sides.

\begin{example}
\label{ex2}
Let $\rho_a=[\psi]$ be unbiased w.r.t.\ both the $v$ and $w$ bases. Then $-\log r(v,w;[\psi])=2\log d$, which is much stronger than the bound $-\log r(v,w)\leq \log d$. Likewise, \eqref{eqn15} gives $ H_{}(v|b)\geq\log d$ whereas \eqref{eqn13} gives $ H_{}(v|b)\geq 0$. A similar strengthening of the bounds occurs if $\rho_a$ is \emph{approximately} unbiased w.r.t.\ $v$ and $w$. Thus the state-dependent bounds account for the complementarity between the state and the measurement(s) of interest. 
\end{example}

\begin{example}
\label{ex3}
Consider a qutrit ($d=3$) with bases $v=\{\ket{0},\ket{1},\ket{2}\}$ and $w=\{\ket{0},\ket{1}+\ket{2},\ket{1}-\ket{2}\}$. Since $r(v,w)=1$, \eqref{eqn10} gives a trivial bound. But if $\rho_a$ lives only in the space spanned by $\ket{1}$ and $\ket{2}$, then set $\Pi=[1]+[2]$, and obtain: $H_{}(v|b)+ H_{}(w|c)\geq \log 2$. This \textit{reveals the hidden complementarity} between $v$ and $w$.
\end{example}

\textit{Minimum uncertainty states}. Because the UPQSI is intimately connected to the monotonicity of the relative entropy, states that satisfy the former with equality are precisely states that satisfy the later with equality. Petz showed \cite{Petz2003,HaydenEtAl04} that $S(\rho||\sg)=S(\EC(\rho)||\EC(\sg))$ if and only if there exists a quantum channel $\hat \EC$ that undoes the action of $\EC$ on $\rho$ and $\sg$:
\begin{equation}
\label{eqn17}
\hat\EC\EC\rho=\rho,\quad \hat\EC\EC \sg = \sg.
\end{equation} 
The construction given for this is \cite{HaydenEtAl04}:
\begin{equation}
\label{eqn18}
\hat\EC(\rho)=\sqrt{\sg}\EC\ad(\EC(\sg)^{-1/2}\rho\EC(\sg)^{-1/2})\sqrt{\sg},
\end{equation} 
which automatically satisfies $\hat\EC\EC \sg = \sg$, so one just needs to solve $\hat\EC\EC\rho=\rho$. We take this approach to finding the MUS for particular uncertainty relations.

In what follows we consider a special pair of MUBs, the $x$ and $z$ bases, which are related by the Fourier transform:
\begin{equation}
\label{eqn19}
\ket{z_k}=\sum_j \frac{\om^{jk}}{\sqrt{d}}\ket{x_j},\quad \ket{x_j}=\sum_k \frac{\om^{-jk}}{\sqrt{d}}\ket{z_k},
\end{equation} 
where $\om=e^{2\pi i/d}$. Consider the uncertainty relations \cite{MaassenUffink, RenesBoileau, BertaEtAl, ColesEtAlv4}:
\begin{align}
\label{eqn20}H(x)+H(z)&\geq \log d,\\
\label{eqn21}H(x)+H(z)&\geq \log d + S(\rho_a),\\
\label{eqn22}H(x)+H(z|b)&\geq \log d + S(a|b),\\
\label{eqn23}H(x|b)+H(z|b)&\geq \log d + S(a|b),
\end{align} 
which are shown in order of increasing generality; \eqref{eqn21} becomes \eqref{eqn20} for unipartite pure states, \eqref{eqn22} becomes \eqref{eqn21} for bipartite product states $\rho_{ab}=\rho_a\ot \rho_b$, and \eqref{eqn23} becomes \eqref{eqn22} for states with $\chi(x,b)=0$. We have found all MUS associated with \eqref{eqn20}, \eqref{eqn21}, and \eqref{eqn22}, and we discuss the MUS for \eqref{eqn23} \footnotemark[\value{footnote}].

\begin{theorem}
\label{thm2}
Let $d$ be prime, then

(i) A state $\rho_a$ is a MUS of \eqref{eqn20} if and only if it is (pure and) a basis state from either the $z$ or $x$ basis.

(ii) A state $\rho_a$ is a MUS of \eqref{eqn21} if and only if it is diagonal in either the $z$ or $x$ basis.

(ii) A state $\rho_{ab}$ is a MUS of \eqref{eqn22} if and only if $\rho_{ab}=\sum_k [z_k]\rho_{ab}[z_k]$ or $\rho_{ab}=\sum_j [x_j]\rho_{a}[x_j]\ot \rho_b$. \openbox 
\end{theorem}

As a corollary to Theorem~\ref{thm2}, we have found the MUSs of the uncertainty relation \cite{ColesEtAlv4} for a qubit ($d=2$):
\begin{equation}
\label{eqn24}
H(x)+H(y)+H(z)\geq 2\log 2 + S(\rho_a),
\end{equation} 
where $x$, $y$, and $z$ are any complete set of three MUBs of the qubit.

\begin{corollary}
\label{thm3}
A state $\rho_a$ is a MUS of \eqref{eqn24} if and only if it is diagonal in either the $x$, $y$, or $z$ basis. \openbox
\end{corollary}

We now generalize Theorem~\ref{thm2} to arbitrary $d$, letting $\{s_\al\}_{\al=1}^\eta$ be the set of all factors of $d$, e.g.\ $\{1,2,4\}$ for $d=4$. It is helpful to introduce the states:
\begin{align}
\label{eqn25}
\ket{w^ \al_{\bt,\gm}}&=\sum_{n=0}^{s_\al-1} \frac{\om^{-n\gm d/s_\al}}{\sqrt{s_\al}}\ket{z_{\bt+nd/s_\al}}\nonumber\\
& =\sum_{m=0}^{d/s_\al-1} \frac{\om^{m \bt s_\al}}{\sqrt{d/s_\al}}\ket{x_{\gm+ms_\al}},
\end{align} 
where $\al=1,...,\eta$; $\bt=0,...,(d/s_\al)-1$; and $\gm=0,...,s_\al-1$. For a fixed $\al$, the set of $\ket{w^ \al_{\bt,\gm}}$ with different $\bt, \gm$ form an orthonormal basis, denoted the $w^\al$ basis. It is sometimes helpful to think of $w^\al$ as a tensor product of the $z$ and $x$ bases respectively on subsystems $a_1$ and $a_2$ of dimension $d/s_\al$ and $s_\al$, i.e.\ $\ket{w^\al_{\bt,\gm}}=\ket{z_\bt}_{a_1}\ket{x_\gm}_{a_2}$. It will also be useful to introduce $p_j=\Tr([x_j]\rho_a)$, $q_k=\Tr([z_k]\rho_a)$, $\sg^x_{b,j}=\Tr_a([x_j]\rho_{ab})$, and $\sg^z_{b,k}=\Tr_a([z_k]\rho_{ab})$.

\begin{theorem}
\label{thm4}
Let $d$ be arbitrary (with $\eta$ factors), then

(i) A state $\rho_a$ is a MUS of \eqref{eqn20} if and only if it is one of the pure states $\ket{w^ \al_{\bt,\gm}}$ given in \eqref{eqn25}, i.e.\ a basis state from one of the $w^\al$ bases.

(ii) The MUS of \eqref{eqn21} are diagonal in one of the $w^\al$ bases, with further constraints on the diagonal elements as follows. Let $\rho_a^\al$ denote the general solution that is diagonal in the $w^\al$ basis, then $\rho_a^\al=d\sum_{\bt,\gm} p_\gm q_\bt [w^\al_{\bt,\gm}]=d(\sum_\bt q_\bt [z_\bt]_{a_1})\ot (\sum_\gm p_\gm [x_\gm]_{a_2})$.

(iii) The MUS of \eqref{eqn22} are $\rho^\al_{ab}=d\sum_{\bt,\gm} p_\gm [w^\al_{\bt,\gm}]\ot \sg^z_{b,\bt}=d(\sum_{\bt} [z_\bt]_{a_1}\ot \sg^z_{b,\bt})\ot (\sum_\gm p_\gm [x_\gm]_{a_2}) $. \openbox
\end{theorem}

Our approach should work for other MUBs as well. For example, the following result for tensor products of $x$ and of $z$ implies Theorem~\ref{thm4} by setting all but one $d_\nu$ to 1.
\begin{theorem}
\label{thm5}
Let $a$ consist of $\lm$ subsystems with $d=d_1...d_\nu...d_\lm$ such that all $d_\nu$ are pairwise coprime, with $\{s^{(\nu)}_{\al_\nu}\}_{\al_\nu=1}^{\eta_\nu}$ the set of factors of $d_\nu$. Then the MUS of
\begin{equation}
\label{eqn26}
H(\bigotimes_{\nu=1}^\lm x_\nu)+ H(\bigotimes_{\nu=1}^\lm z_\nu|b)\geq \log d +S(a|b)
\end{equation}
have the form $\rho^{\vec \al}_{ab}=d\sum_{\vec\bt, \vec\gm} p_{\vec\gm} (\bigotimes_{\nu=1}^\lm[w^{\al_\nu}_{\bt_\nu,\gm_\nu}])\ot \sg^z_{b, \vec\bt}$, where $\vec\al=(\al_1,...,\al_\lm)$ and likewise for $\vec\bt$ and $\vec\gm$, with $\bt_\nu=0,...,d_\nu/s^{(\nu)}_{\al_\nu}-1$ and $\gm_\nu=0,...,s^{(\nu)}_{\al_\nu}-1$. \openbox
\end{theorem}

\textit{MUS of \eqref{eqn23}}. The MUS of \eqref{eqn23} are tripartite pure states $\rho_{abc}$ that satisfy
\begin{equation}
\label{eqn27}
 H_{}(x|b)+ H_{}(z|c)= H_{}(x|c)+ H_{}(z|b)= \log d.
\end{equation} 
Let us denote with $\Xi$ the set of all states for which at least one of the four $ H_{}(\cdot |\cdot )$ terms in \eqref{eqn27} is zero. Renes and Boileau \cite{RenesBoileau} noted that all states in $\Xi $ satisfy \eqref{eqn27} and remarked that it is an open question as to whether $\Xi$ are the \emph{only} states that satisfy \eqref{eqn27}. Theorem~\ref{thm4} shows that there are other solutions in non-prime $d$, e.g.\ the states in \eqref{eqn25} satisfy \eqref{eqn27} with $ H_{}(z|b)= H_{}(z|c)=H(z)=\log s_\al$ and $ H_{}(x|b)= H_{}(x|c)=H(x)= \log (d/s_\al)$. Generally, instead of just four solutions (as in $\Xi $), one should consider $2\eta$ solutions that, for some $\al$, have either $H(w^\al |c)=0$ or $H(w^\al |b)=0$, with further constraints given in \footnotemark[\value{footnote}]; denote this set of MUS as $\Upsilon $, so $\Upsilon \supseteq \Xi$. 

However, there is an entirely different \emph{class}, $\Omega$, of states that satisfy \eqref{eqn27}. Consider the tripartite state with $0<g<1$: $\ket{\psi}_{abc}=\sqrt{g}\ket{x_j}\ket{0}\ket{0}+\sqrt{1-g}\ket{z_k}\ket{1}\ket{1}$, for which $\rho_{ab}=\rho_{ac}=g[x_j]\ot [0]+(1-g) [z_k]\ot [1]$. Since $H(x|b)=H(x|c)=(1-g)\log d$ and $H(z|b)=H(z|c)=g\log d$, this is a solution to \eqref{eqn27} that is not in $\Upsilon$. (Note that this sort of MUS works for arbitrary MUBs, not just $x$ and $z$.) More generally, $\Om$ contains:
\begin{equation}
\label{eqn28}
\rho_{ab}=\sum_{\al,\bt,\gm} g_{\al,\bt,\gm} [w^\al_{\bt,\gm}]\ot \rho_{\al,\bt,\gm},
\end{equation} 
where the different $\rho_{\al,\bt,\gm}$ are all orthogonal and $0\leq g_{\al,\bt,\gm}\leq 1$. 

Finally, we believe there is a third class of MUS, $\Lm$, that is neither in $\Upsilon$ nor $\Omega$. For example in $d=2$, any state of the form $\ket{\psi}_{abc}=(\ket{0}\ket{\phi_b}\ket{\phi_c}+\ket{1}\ket{\varphi_b}\ket{\varphi_c})/\sqrt{2}$, where $\ket{\phi_b},\ket{\phi_c},\ket{\varphi_b},\ket{\varphi_c}$ are arbitrary kets with $\ip{\phi_b}{\varphi_b}\ip{\phi_c}{\varphi_c}\in \mathbb{R}$, satisfies \eqref{eqn27} with $H(z|b)=\log 2 - S(\rho_b)$ and $H(x|c)= S(\rho_b)$. The three classes are seen as distinct as follows: in $\Upsilon$, either $\rho_{ab}$ or $\rho_{ac}$ has zero discord \cite{OllZur01}; in $\Omega$, $\rho_{ab}$ and $\rho_{ac}$ are separable with non-zero discord; in $\Lm$, $\rho_{ab}$ and $\rho_{ac}$ are entangled \footnotemark[\value{footnote}].

Berta et al. \cite{BertaEtAl} outlined methods for using the UPQSI \eqref{eqn10} for witnessing entanglement and for quantum cryptography. For both applications, one essentially lower-bounds the entanglement of $\rho_{ab}$ with, e.g.\ $-S(a|b)\geq \log d - H(x|b)-H(z|b)$, where Alice and Bob find upper-bounds: $H(x|b)\leq H(x|x)$ and $H(z|b)\leq H(z|z)$ by comparing their measurement results in the $x$ and $z$ bases on an unknown state $\rho_{ab}$. The MUS are precisely the states for which this method should work best (otherwise the bound on the entanglement would be loose); providing motivation for further studying MUS. 

In summary, the entropic uncertainty principle can be viewed as a data-processing inequality, expressing the notion that information cannot increase in the process shown in Fig.~\ref{fgr1}. Finding minimum uncertainty states then maps onto the question of whether this process is reversible, or whether information is irreversibly lost. 

We thank Robert Griffiths for helpful conversations. This research is supported by the Office of Naval Research and the U.S. Department of Energy through the LANL/LDRD Program.

\bibliography{RelEntropyProof}

\section{Supplemental Material}

Here we give the proofs of Theorems~1 through 5 of the main manuscript, and also elaborate on the MUS of (23). (We preserve the numbering of the equations and theorems in the main manuscript, and add a prefix ``S" to such objects appearing in this supplemental material.) Let us first state the following useful result, proved in \cite{ColesEtAlv4}, that relates the conditional entropy to the relative entropy. This will allow us to rewrite the UPQSI in terms of relative entropy.

\begin{thm4}
\label{thm6}
Let $\Pi=\{\Pi_{j}\}$ be a projective decomposition of $I_a$ and let $P=\{P_{j}\}$ be a POVM on $a$.

(i) Let $\rho_{abc}$ be a pure state, then
\begin{equation}
\label{eqn29}
\tag{S1}
 H_{}(\Pi|b)=S(\rho_{ac}||\sum_j \Pi_{j}\rho_{ac}\Pi_{j}).
\end{equation} 

(iii) Let $\rho_{abc}$ be \emph{any} state, then
\begin{equation}
\tag{S2}
\label{eqn30}
 H_{}(P|b)\geq S(\rho_{ac}||\sum_j P_{j}\rho_{ac}P_{j}).
\end{equation} \openbox
\end{thm4}

\subsection{Proof of Theorem~1}

First, consider the single-POVM UPQSI in (15). We remarked in the main manuscript that the strongest bound in (15) results from chosing $\Pi$ to have the smallest possible rank, i.e.\ the projector onto the support of $\rho_a$. One can see this by considering two projectors $\Pi$ and $\Pi'$ where the latter has a higher rank than the former and $\Pi'=\Pi+\Phi$ where $\Phi$ is also a projector, and note that $G'_j=\sqrt{P_{j}}\Pi'\sqrt{P_{j}}\geq \sqrt{P_{j}}\Pi\sqrt{P_{j}}=G_j$. It follows \cite{HornJohn} that the spectrum of $G'_j$ weakly majorizes that of $G_j$ and thus $\|\Pi' \sqrt{P_{j}}\|_\infty^2 \geq \|\Pi \sqrt{P_{j}}\|_\infty^2$. Now let us prove (15).

\begin{proof}
The important properties \cite{VedralReview02, OhPe93} of $S(\cdot||\cdot)$ we use are:
\begin{equation}
\tag{S3}\label{eqn31}
S(\rho||\sg)\geq S(\EC(\rho)||\EC(\sg))
\end{equation} 
for any quantum channel $\EC$; and for positive operators $\rho$, $\sg $, $\tau $, if $\tau \geq \sg$, then
\begin{equation}
\tag{S4}\label{eqn32}
S(\rho||\sg)\geq S(\rho ||\tau).
\end{equation} 
Let $\lm_{\max}(A)$ denote the maximum eigenvalue of $A$, let $G_{j}= \sqrt{P_{j}} \Pi\sqrt{P_{j}} $, note $\lm_{\max}(G_j)=\|\Pi \sqrt{P_{j}}\|_\infty^2$, then from \eqref{eqn30}:
\begin{align}
 H_{}(P|b)&\geq S(\rho_{ac}||\sum_j P_{j}\rho_{ac}P_{j})\nonumber\\
\tag{S5}\label{eqn33}&\geq S(\rho_{ac}||\sum_j \Pi P_{j}\rho_{ac}P_{j} \Pi) \\
\tag{S6}\label{eqn34}&\geq S(\rho_{c}||\sum_j \Tr_a\{\Pi P_{j}\rho_{ac}P_{j} \Pi\}) \\
\tag{S7}\label{eqn35}&\geq S(\rho_{c}||\sum_j \lm_{\max}(G_{j}) \Tr_a\{P_{j}\rho_{ac}\}) \\
\tag{S8}\label{eqn36}&\geq S(\rho_{c}|| \max_j \lm_{\max}(G_{j}) \rho_c])\\
\tag{S9}\label{eqn37}&=-\log \max_j \lm_{\max}(G_{j}).
\end{align}
We invoked \eqref{eqn31} for \eqref{eqn33} with the channel $\rho\to\Pi\rho\Pi+(I-\Pi)\rho(I-\Pi)$, and for \eqref{eqn34} with the channel $\rho\to\Tr_a\rho$. We invoked \eqref{eqn32} for \eqref{eqn35}; $\lm_{\max}(G_{j})I_a\geq G_j$ which implies $\Tr_a[\lm_{\max}(G_{j})I_aT_{ac,j}]\geq \Tr_a[G_jT_{ac,j}]$, where $T_{ac,j}=\sqrt{P_j}\rho_{ac}\sqrt{P_j}$ is a positive operator. We also used \eqref{eqn32} for \eqref{eqn36}, i.e.\ $\max_j \lm_{\max}(G_{j}) \sum_j A_j \geq \sum_j \lm_{\max}(G_{j})A_j$ where the $A_j$ are positive operators.

\end{proof}

Now we prove (14).
\begin{proof}
Let $e$ be an auxiliary system that acts as a register for the $Q$ measurement. Consider the quantum channel $\EC_ Q \colo ab\rightarrow eb$ defined by $\EC_ Q(\rho_{ab})=\sum_k [e_k]\ot \Tr_a(Q_{k}\rho_{ab})$, where $\{\ket{e_k}\}$ is an orthonormal basis of $e$.  Also, define $G_{jk}= \sqrt{P_{j}} \Pi Q_{k} \Pi\sqrt{P_{j}} $, and note $G_{jk} \leq \lm_{\max}(G_{jk})I_a$, and $r(P,Q; \Pi)=\max_{j,k}\lm_{\max}(G_{jk})$. Then, starting from \eqref{eqn33} (swapping labels $b$ and $c$), 

\begin{align}
\tag{S10}\label{eqn38}& H_{}(P|c)\geq S(\rho_{ab}||\sum_j \Pi P_{j}\rho_{ab}P_{j}\Pi)\\
\tag{S11}\label{eqn39}&\geq S(\EC_ Q(\rho_{ab})||\sum_j \EC_ Q(\Pi P_{j}\rho_{ab}P_{j}\Pi))\\
&= S(\sum_{k} [e_k]\ot \Tr_a\{Q_{k} \rho_{ab}\}||\nonumber\\
\tag{S12}\label{eqn40}&\hspace{3pt}\sum_{j,k} [e_k]\ot \Tr_a\{G_{jk}\sqrt{P_j}\rho_{ab}\sqrt{P_j}\})\\
&\geq S(\sum_{k} [e_k]\ot \Tr_a\{Q_{k} \rho_{ab}\}||\nonumber\\
\tag{S13}\label{eqn41}& \hspace{3pt} \sum_{j,k} \lm_{\max}(G_{jk})[e_k]\ot \Tr_a\{P_{j}\rho_{ab}\})\\
\tag{S14}\label{eqn42}&\geq S(\sum_{k} [e_k]\ot \Tr_a\{Q_{k} \rho_{ab}\}|| r(P,Q; \Pi) I_e \ot \rho_{b})\\
\tag{S15}\label{eqn45}&= -\log  r(P,Q; \Pi) -  H_{}(Q|b),
\end{align}
We invoked \eqref{eqn31} for step \eqref{eqn39}, \eqref{eqn32} for steps \eqref{eqn41} and \eqref{eqn42}, and Eq.~(11.58) of \cite{NieChu00} for step \eqref{eqn45}.
\end{proof}

\subsection{Proof of Theorem~2}

\begin{proof}
This theorem can be viewed as a corollary to Theorem~4. Set $d$ to be prime, so that $\eta=2$ and $\{s_\al\}=\{1,d\}$. For $s_\al=1$, $w^\al$ is the $z$-basis, and for $s_\al=d$, $w^\al$ is the $x$-basis. Thus, part (i) of Theorem~4 clearly reduces to part (i) of Theorem~2. Part (ii) of Theorem~4 reduces to part (ii) of Theorem~2 since there are no constraints on the diagonal elements of $\rho^\al_a$ for $s_\al$ equal to 1 or $d$. Likewise, setting $s_\al$ equal to 1 or $d$ in $\rho^\al_{ab}=d\sum_{\bt,\gm} p_\gm [w^\al_{\bt,\gm}]\ot \sg^z_{b,\bt}$ gives the two solutions in part (iii) of Theorem~2. 
\end{proof}

\subsection{Proof of Corollary~3}

\begin{proof}
Define $\zeta:=H(x)+H(y)+H(z)- 2\log 2 - S(\rho_a)$. First, consider (possibly mixed) states $\rho_a$ in the $xy$ plane of the Bloch sphere; such states have $H(z)=\log 2$. For these states, $\zeta =0$ if and only if $H(x)+H(y)=\log 2 +S(\rho_a)$. But from Theorem~2, this is true if and only if either $x$ or $y$ is the eigenbasis of $\rho_a$, i.e. the state lies on either the $x$ or $y$ axis of the Bloch sphere. Any other state in the $xy$ plane will strictly have $H(x)+H(y)>\log 2 +S(\rho_a)$. Now consider taking a vertical path in the Bloch sphere up from some point in the $xy$ plane. Such a path will never decrease the value of $\zeta$ (See Appendix F of \cite{ColesEtAlv4}). Thus, the only states that could possibly satisfy $\zeta =0$ are those in the $xz$ plane and the $yz$ plane. But we already know that the territory between the $x$ and $y$ axes in the $xy$ plane cannot have $\zeta=0$, so by symmetry, the territory between the $x$ and $z$ axes in the $xz$ plane cannot have $\zeta=0$, and likewise for the $yz$ plane. So the only states that satisfy $\zeta =0$ are those along the $x$, $y$, and $z$ axes.
\end{proof}

\subsection{Proof of Theorem~4}

\begin{proof}
Even though this is a corollary of Theorem~5, it is instructive to see the direct proof as it is simpler than that of Theorem~5. We discuss below that parts (i) and (ii) follow from part (iii).

(i) Clearly from (21) the only states that can satisfy (20) with equality are pure states $[S(\rho_a)=0]$. Thus, the MUS of (20) are a subset of the MUS of (21), precisely the subset with $S(\rho_a)=0$. Assuming part (ii) of this theorem is true, then the only states that can be MUS of (21) are diagonal in a $w^\al$ basis, and thus the only states that can be MUS of (20) are (pure) basis vectors from a $w^\al$ basis, and indeed it is easily verified that all such basis vectors are MUS of (20).

(ii) Likewise part (ii) follows from part (iii) of this theorem. The MUS of (21) are a subset of the MUS of (22), precisely the subset with $\rho_{ab}=\rho_a \ot \rho_b$. Imposing this condition on $\rho^\al_{ab}=d\sum_{\bt,\gm} p_\gm [w^\al_{\bt,\gm}]\ot \sg^z_{b,\bt}$ and tracing over $b$ gives $\rho^\al_{a}=d\sum_{\bt,\gm} p_\gm q_\bt [w^\al_{\bt,\gm}]$. (It turns out we did not need to impose the condition $\rho_{ab}=\rho_a\ot\rho_b$ since all MUS of (22) have a $\rho^\al_a$ of this form.)

(iii) It remains only to prove part (iii). Using (17) and (18) with $\rho=\rho_{ab}$, $\sg= \sum_j [x_{j}] \rho_{ab}[x_{j}]$, $\EC(\cdot)=\sum_k [z_{k}](\cdot)[z_{k}]= \EC\ad(\cdot)$, gives:
\begin{equation}
\tag{S16}\label{eqn46}
 \rho_{ab}=\sum_{j,j',k} \om^{(j-j')k} \dyad{x_j}{x_{j'}}\ot \sqrt{\sg^x_{b,j}}\rho_b^{-1/2} \sg^z_{b,k}\rho_b^{-1/2} \sqrt{\sg^x_{b,j'}}.
\end{equation} 
Now specializing to $\chi(x,b)=0$, meaning $\sg^x_{b,j}=p_j\rho_b$ for each $j$, \eqref{eqn46} becomes:
\begin{equation}
\tag{S17}\label{eqn47}
\rho_{ab}= \sum_{j,j'} \sqrt{p_jp_{j'}} \dyad{x_j}{x_{j'}}\ot \Tr_a(Z^{j-j'}\rho_{ab}),
\end{equation} 
where $Z=\sum_k \om^k [z_k]$. Computing $\Tr_a(Z^\mu\rho_{ab})$ from \eqref{eqn47} for $\mu=1,..., d-1$, one arrives at a system of equations (one for each $\mu$):
\begin{equation}
\tag{S18}\label{eqn48}f_\mu(z)g_\mu(x)=0,
\end{equation} 
where
\begin{align}
f_\mu(z)&:= \Tr_a(Z^\mu\rho_{ab})=\sum_k \om^{\mu k} \sg^z_{b,k},\nonumber\\
\tag{S19}\label{eqn49} g_\mu(x)&:=1-\sum_{j} \sqrt{p_jp_{j+\mu}}.
\end{align} 

One can show that $g_\mu(x)=0$ if and only if $p_j=p_{j+m\mu}$ for all $j,m\in \mathbb{Z}_d$, as follows. Using the method of Lagrange multipliers, the Lagrangian is $L=1-\sum_{j} \sqrt{p_jp_{j+\mu}}+\lm(1-\sum_j p_j)$. Taking $\partial L/\partial p_k=0$ gives $-2\lm \sqrt{p_k}= \sqrt{p_{k+\mu}}+ \sqrt{p_{k-\mu}}$, and summing this over all $k$ shows that $\lm=-1$. Thus rearranging: $ \sqrt{p_k}-\sqrt{p_{k-\mu}}= \sqrt{p_{k+\mu}}- \sqrt{p_k}$, which must also equal $\sqrt{p_{k+2\mu}}- \sqrt{p_{k+\mu}}$, etc. Since each stepwise difference is the same and doing $d$ steps brings us back to the same point ($p_k=p_{k+d\mu}$), it must be that $p_k=p_{k+m\mu}$ for all $m=0,...,d-1$.

Now note that $g_\mu(x)=0$ implies that $g_{m\mu}(x)=0$. This fact implies that there are $\eta$ and only $\eta$ different ways to set some $g_{\mu}(x)$ terms to zero, each way corresponding to setting $g_{s_\al}(x)=g_{ms_\al}(x)=0$, thus $p_j=p_{j+m s_\al}$ for $m=0,...,(d/s_\al)-1$,  and $g_\mu(x)\neq 0$ for $\mu\neq ms_\al$. Of course, to solve the system of equations, \eqref{eqn48}, one must compensate for the non-zero $g_\mu(x)$ by setting $f_\mu(z)=0$ for $\mu\neq ms_\al$, which can be shown to imply that $\sg^z_{b,k}=\sg^z_{b,k+nd/s_\al}$ for $n=0,...,s_\al-1$, as follows. Noting that $\mu$ and $k$ are Fourier partners, Fourier-transform $f_\mu(z)$ to get $\sg^z_{b,k}=(1/d)\sum_{\mu} \om^{-\mu k}f_\mu(z)=(1/d)\sum_{m} \om^{-m s_\al k}f_{m s_\al}(z)$. Clearly this implies that $\sg^z_{b,k}=\sg^z_{b,k+nd/s_\al}$.

Thus we have $\eta$ solutions where the $\al$-th solution, denoted $\rho^\al_{ab}$, has the properties that $p_j=p_{j+m s_\al}$ and $\sg^z_{b,k} = \sg^z_{b,k+nd/s_\al}$. Now we rewrite the $\rho_{ab}$ in \eqref{eqn47}, letting $j=\gm+m s_\al$, $j'=\gm'+m' s_\al$, $k=\bt+nd/s_\al$, with $0\leq \gm,\gm',n \leq s_\al-1$ and $0\leq \bt,m,m' \leq d/s_\al-1$, giving:
\begin{align}
&\rho_{ab}= \sum_{\gm,\gm',m,m',\bt,n}  \om^{(\bt+nd/s_\al)(\gm-\gm'+ms_\al-m's_\al)}\times\nonumber\\
\tag{S20}\label{eqn50}&\sqrt{p_{\gm+ms_\al} p_{\gm'+m's_\al}}\dyad{x_{\gm+ms_\al}}{x_{\gm'+m's_\al}}\ot \sg^z_{b,\bt+nd/s_\al}.
\end{align} 
So for the $\al$-th solution this reduces to:
\begin{align}
\rho^\al_{ab}= & \sum_{\gm,\gm',m,m',\bt,n}  \om^{(\bt+nd/s_\al)(\gm-\gm'+ms_\al-m's_\al)}\times\nonumber\\
\tag{S21}\label{eqn51}&\hspace{26pt} \sqrt{p_\gm p_{\gm'}}\dyad{x_{\gm+ms_\al}}{x_{\gm'+m's_\al}}\ot \sg^z_{b,\bt},
\end{align} 
The sum over $n$ gives a $\dl_{\gm,\gm'}$ and we arrive at:
\begin{align}
\rho^\al_{ab}&=  s_\al \sum_{\gm,m,m',\bt}  \om^{\bt(ms_\al-m's_\al)}\times\nonumber\\
\tag{S22}\label{eqn52}&\hspace{31pt} p_\gm \dyad{x_{\gm+ms_\al}}{x_{\gm+m's_\al}}\ot \sg^z_{b,\bt},
\end{align} 
Using $\sqrt{d}\ket{w^\al_{\bt,\gm}}= \sqrt{s_\al} \sum_m \om^{\bt ms_\al} \ket{x_{\gm+ms_\al}}$, we arrive at $\rho^\al_{ab}=d\sum_{\bt,\gm}p_\gm [w^\al_{\bt,\gm}]\ot \sg^z_{b,\bt}$.
\end{proof}

\subsection{Proof of Theorem~5}

\begin{proof}
This proof mirrors that of Theorem~4, except now we use a vector notation for all quantities, e.g.\ $\vec j=(j_1,...,j_\lm)$ and $\vec \mu=(\mu_1,...,\mu_\lm)$, where each component refers to a particular subsystem. From (17) and (18), the MUS of (26) are:
\begin{equation}
\tag{S23}\label{eqn53}
 \rho_{ab}=\sum_{\vec j,\vec j'} \sqrt{p_{\vec j}p_{\vec j'}} (\bigotimes_{\nu=1}^\lm \dyad{x_{j_\nu}}{x_{j'_\nu}})\ot \Tr_a\{(\bigotimes_{\nu=1}^\lm Z_\nu^{j_\nu-j'_\nu})\rho_{ab}\},
\end{equation} 
where $Z_\nu=\sum_{k_\nu}\om_\nu^{k_\nu}[z_{k_\nu}]$ and $\om_\nu=e^{2\pi i/ d_\nu}$. Now let $\mu_\nu=0,...,d_\nu-1$, compute $\Tr_{a}\{(\bigotimes_{\nu=1}^\lm Z_\nu^{\mu_\nu})\rho_{ab}\}$ and using $\Tr_{a_\nu}(Z_\nu^{\mu_\nu} \dyad{x_{j_\nu}}{x_{j'_\nu}})=\dl_{j_\nu,j'_\nu+\mu_\nu}$ arrive at a system of equations:
\begin{equation}
\tag{S24}\label{eqn54}
 f_{\vec\mu} (z) g_{\vec\mu} (x)=0
\end{equation} 
where
\begin{align}
f_{\vec\mu}(z)&:= \Tr_{a}\{(\bigotimes_{\nu=1}^\lm Z_\nu^{\mu_\nu})\rho_{ab}\} =\sum_{\vec k} (\prod_{\nu=1}^\lm \om_\nu^{\mu_\nu k_\nu}) \sg^z_{b,\vec k},\nonumber\\
\tag{S25}\label{eqn55}
g_{\vec\mu}(x)&:=1-\sum_{\vec j} \sqrt{p_{\vec j}p_{\vec j+\vec\mu}}.
\end{align} 
Consider the following rules. Rule (1): $g_{\vec\mu}(x)=0$ if and only if $p_{\vec j}=p_{\vec j+ \vec \mu}$ for all $j$. This implies the following rules. Rule (2): If $g_{\vec\mu}(x)=0$ then $g_{m \vec\mu}(x)=0$ for all $m=0,...,d-1$, where $m \vec\mu= \vec\mu+ \vec\mu+...$ ($m$ times). Rule (3): If the $d_\nu$ are pairwise coprime and if $g_{\vec\mu}(x)=0$ then $g_{\vec m \vec\mu}(x)=0$ for all $\vec m=(m_1,...,m_\lm)$, where $m_\nu=0,...,d_\nu-1$ and $\vec m \vec\mu=(m_1\mu_1,...,m_\lm \mu_\lm)$. 

Rule (1) follows by the method of Lagrange multipliers, as in the proof of Theorem~4. Rule (2) follows from Rule (1) in a straightforward way. Rule (3) follows from Rule (2) by the Chinese Remainder Theorem, which implies that the ring $\mathbb{Z}_d$ is isomorphic to the ring $\mathbb{Z}_{d_1}\times ...\times\mathbb{Z}_{d_\lm}$. The bijection relating $m\in \mathbb{Z}_d$ to $\vec m\in \mathbb{Z}_{d_1}\times ...\times\mathbb{Z}_{d_\lm}$ is defined through $m_\nu=(m \mod d_\nu)$. By this bijection and the ring isomorphism, $\{g_{m \vec\mu}(x); m=0,..., d-1\}=\{g_{\vec m \vec\mu}(x); m_\nu=0,..., d_\nu-1\}$, and so Rule (3) follows.

From the above rules and letting $\{s^{(\nu)}_{\al_\nu}\}_{\al_\nu=1}^{\eta_\nu}$ be the set of factors of $d_\nu$, there are only $\prod_{\nu=1}^\lm \eta_\nu$ different ways to set some of the $g_{\vec\mu}(x)$ terms to zero, one for each $\vec \al$. The way corresponding to a particular $\vec \al$ involves setting $g_{\vec s_\al}(x)=g_{\vec m \vec s_\al}(x)=0, \forall \vec m$, where $\vec s_\al=(s^{(1)}_{\al_1},..., s^{(\lm)}_{\al_\lm})$, and $g_{\vec \mu}(x)\neq 0$ for $\vec \mu \neq \vec m \vec s_\al$. Of course, to solve \eqref{eqn54} we must set $f_{\vec\mu}(z)=0$ for $\vec \mu \neq \vec m \vec s_\al$.  From the latter condition, it follows that $\sg^z_{b,\vec k}= \sg^z_{b,\vec k+\vec n \vec t_\al}, \forall \vec n$, where $\vec t_\al=(d_1/s^{(1)}_{\al_1},..., d_\lm/s^{(\lm)}_{\al_\lm})$. And from Rule (1), we have $p_{\vec j}=p_{\vec j+ \vec m \vec s_\al}, \forall \vec m$. Plug these two conditions into \eqref{eqn53}, make the variable changes (like in the proof of Theorem~4) $\vec{j}=\vec{\gm}+ \vec m \vec s_\al$, $\vec j'=\vec \gm'+ \vec m' \vec s_\al$, and $\vec{k}=\vec \bt+ \vec n \vec t_\al$, then sum over $\vec n$ to get a $\dl_{\vec \gm,\vec \gm'}$, then change the $x_\nu$ bases to the $w^{\al_\nu}$ bases to arrive at $\rho^{\vec \al}_{ab}=d\sum_{\vec\bt, \vec\gm} p_{\vec\gm} (\bigotimes_{\nu=1}^\lm[w^{\al_\nu}_{\bt_\nu,\gm_\nu}])\ot \sg^z_{b, \vec\bt}$.

\end{proof}

\subsection{MUS of (23)}

Here we discuss different classes of MUS of (23). We remind that reader that discord is a measure of the non-classicality of bipartite correlations. All of our discussion will refer to the one-way discord, as originally defined in \cite{OllZur01}, that is asymmetric under interchanging the two systems; in particular, the discord that uses projectors on system $a$.

Generally, any bipartite state can be classified as either zero-discord (ZD), separable with non-zero discord (SNZD), or entangled (E) \cite{OllZur01}. We shall classify MUS of (23) by classifying the reduced density operators $\rho_{ab}$ and $\rho_{ac}$ of the tripartite pure state $\rho_{abc}$ into one of these three categories, i.e.\ by giving an ordered pair of form ($\rho_{ab}$ category, $\rho_{ac}$ category), for example (ZD,E) means $\rho_{ab}$ is ZD and $\rho_{ac}$ is E. Naively this would give $3\times 3 = 9$ possible ordered pairs, but if $\rho_{ab}$ is ZD then $\rho_{ac}$ cannot be SNZD, and vice-versa. (The proof for this is as follows: If $\rho_{ab}$ is ZD, then there exists a basis $w$ for which $H(w|c)=0$. In turn, if $H(w|b)=0$ then $\rho_{ac}$ is ZD, otherwise if $H(w|b)>0$ then $H(w|c)-H(w|b)=S(a|c)<0$ implying that $\rho_{ac}$ is E. So the only possibilities are for $\rho_{ac}$ to be ZD or E, it cannot be SNZD.) So there are only seven possible ordered pairs, and all seven are physically possible.

Below we find three classes of MUS of (23): one class denoted $\Lm$ for which both $\rho_{ab}$ and $\rho_{ac}$ are E, so (E,E); one class denoted $\Omega$ for which both $\rho_{ab}$ and $\rho_{ac}$ are SNZD, so (SNZD,SNZD); and one class denoted $\Upsilon$ where either $\rho_{ab}$ or $\rho_{ac}$ are ZD, so this includes three ordered pairs (ZD,ZD), (ZD,E), and (E,ZD). It remains an open question as to whether there are MUS of (23) of the form (SNZD,E) or (E,SNZD).

From (17) and (18), the MUS of (23) are tripartite pure states $\rho_{abc}$ with:
\begin{equation}
\tag{S26}\label{eqn56}
 \rho_{ab}=\sum_{j,j',k} \om^{(j-j')k} \dyad{x_j}{x_{j'}}\ot \sqrt{\sg^x_{b,j}}\rho_b^{-1/2} \sg^z_{b,k}\rho_b^{-1/2} \sqrt{\sg^x_{b,j'}}
\end{equation} 
and by symmetry the MUS also satisfy an equation analogous to \eqref{eqn56} for $\rho_{ac}$.

Let us consider solutions $\rho^\al_{abc}$ with the properties that $\sg^x_{b,\gm}=\sg^x_{b,\gm+n s_\al}$ and $\sg^z_{b,\bt}=\sg^z_{b,\bt+m d/s_\al}$ for all $n=0,...,d/s_\al-1$ and all $m=0,...,s_\al-1$; and other solutions $\rho^{\eta+\al}_{abc}$ with $\sg^x_{c,\gm}=\sg^x_{c,\gm+n s_\al}$ and $\sg^z_{c,\bt}=\sg^z_{c,\bt+m d/s_\al}$ likewise for all $n$ and $m$. Then from \eqref{eqn56}:
\begin{align}
\tag{S27}\label{eqn57}
 \rho^\al_{ab}&=d \sum_{\bt,\gm} [w^\al_{\bt\gm}] \ot A_{b;\bt,\gm}\ad A_{b;\bt,\gm},\\
 \tag{S28}\label{eqn58}
 \rho^{\eta+\al}_{ac}&=d \sum_{\bt,\gm} [w^\al_{\bt\gm}] \ot A_{c;\bt,\gm}\ad A_{c;\bt,\gm},
\end{align} 
where $A_{b;\bt,\gm}= \sqrt{\sg^z_{b,\bt}} \rho_b^{-1/2} \sqrt{\sg^x_{b,\gm}} $ and $A_{c;\bt,\gm}= \sqrt{\sg^z_{c,\bt}} \rho_c^{-1/2} \sqrt{\sg^x_{c,\gm}}$, and as always $\bt=0,...,d/s_\al-1$ and $\gm=0,...,s_\al-1$. Note that the solution $\rho^{\al}_{abc}$ has $H(w^\al |c)=0$, while the solution $\rho^{\eta+\al}_{abc}$ has $H(w^\al |b)=0$. These represent the $2\eta$ solutions ($\eta$ is the number of factors of $d$, e.g.\ $\eta=3$ for $d=4$) described in the main manuscript that compose the set $\Upsilon$. Setting $s_\al=1$ or $s_\al=d$ in \eqref{eqn57} and \eqref{eqn58} shows that $\Upsilon$ contains all states for which either $H(z|c)$, $H(x|c)$, $H(z|b)$, or $H(x|b)$ equals zero, and so $\Upsilon$ contains the set $\Xi$ defined in the main manuscript.

Let us consider a second class $\Om$ of MUS of the form:
\begin{equation}
\tag{S29}\label{eqn59}
\rho_{ab}=\sum_{\al,\bt,\gm} g_{\al,\bt,\gm} [w^\al_{\bt,\gm}]\ot \rho_{\al,\bt,\gm},
\end{equation} 
where the different $\rho_{\al,\bt,\gm}$ are all orthogonal to each other and $0\leq g_{\al,\bt,\gm}\leq 1$. For these states $S(a|b)=0$, $H(z|b)=H(z|c)=\sum g_{\al,\bt,\gm} H(z)_{\ket{w^\al_{\bt,\gm}}}= \sum_{\al,\bt,\gm} g_{\al,\bt,\gm}\log s_\al$, and $H(x|b)=H(x|c)= \sum g_{\al,\bt,\gm} H(x)_{\ket{w^\al_{\bt,\gm}}}=\sum_{\al,\bt,\gm} g_{\al,\bt,\gm}\log (d/ s_\al)$. So they satisfy (27) since $\sum_{\al,\bt,\gm} g_{\al,\bt,\gm}=1$. Also, one can show (with a Schmidt decomposition across the $ab/c$ cut) that if $\rho_{ab}$ is given by \eqref{eqn59}, then $\rho_{ac}$ has the same form:
\begin{equation}
\tag{S30}\label{eqn60}
\rho_{ac}=\sum_{\al,\bt,\gm} g_{\al,\bt,\gm} [w^\al_{\bt,\gm}]\ot \sg_{\al,\bt,\gm},
\end{equation} 
where the different $\sg_{\al,\bt,\gm}$ are all orthogonal to each other. Thus, both $\rho_{ab}$ and $\rho_{ac}$ are separable, and as long as more than one $w^\al$ basis appears in the sums in \eqref{eqn59} and \eqref{eqn60}, then they both have non-zero discord.

Finally, the main manuscript gives an example for $d=2$ of MUS that are neither in $\Upsilon$ nor in $\Omega$. The tripartite state:
\begin{equation}
\tag{S31}\label{eqn61}
\ket{\psi}_{abc}=(\ket{0}\ket{\phi_b}\ket{\phi_c}+\ket{1}\ket{\varphi_b}\ket{\varphi_c})/\sqrt{2}
\end{equation} 
where $\ket{\phi_b},\ket{\phi_c},\ket{\varphi_b},\ket{\varphi_c}$ are arbitrary kets with $\ip{\phi_b}{\varphi_b}\ip{\phi_c}{\varphi_c}\in \mathbb{R}$, satisfies (27) with $H(z|b)=\log 2 - S(\rho_b)$, $H(z|c)=\log 2 - S(\rho_c)$, $H(x|b)=S(\rho_c)$, and $H(x|c)= S(\rho_b)$. Likewise, replacing the $z$ states $\{\ket{0},\ket{1}\}$ in \eqref{eqn61} with the $x$ states $\{\ket{+},\ket{-}\}$, the tripartite state:
\begin{equation}
\tag{S32}\label{eqn62}
\ket{\psi}_{abc}=(\ket{+}\ket{\phi_b}\ket{\phi_c}+\ket{-}\ket{\varphi_b}\ket{\varphi_c})/\sqrt{2}
\end{equation} 
satisfies (27) with $H(x|b)=\log 2 - S(\rho_b)$, $H(x|c)=\log 2 - S(\rho_c)$, $H(z|b)=S(\rho_c)$, and $H(z|c)= S(\rho_b)$. Except for the extreme cases where $S(\rho_b)$ or $S(\rho_c)$ are 0 or $\log 2$, the states described by \eqref{eqn61} and \eqref{eqn62} are clearly not in $\Upsilon$, and the fact that they are not in $\Omega$ follows from $S(b|a)=-S(b|c)<0$ and $S(c|a)=-S(c|b)<0$, implying that both $\rho_{ab}$ and $\rho_{ac}$ are entangled, in contrast to the separable states in $\Omega$. There is reason to believe that there are MUS for $d>2$ of a similar nature to the qubit examples given here (with both $\rho_{ab}$ and $\rho_{ac}$ entangled), as we have found such MUS for $d=3$. For example:  
\begin{equation}
\tag{S33}\label{eqn63}
\ket{\psi}_{abc}=(\ket{z_0}\ket{0}\ket{0}+ \ket{z_1}\ket{+}\ket{+}+ \ket{z_2}\ket{y+}\ket{y-})/\sqrt{3},
\end{equation} 
where $b$ and $c$ are qubits and $\ket{y\pm}=(\ket{0}\pm i \ket{1})/\sqrt{2}$, has $H(z|b)= H(z|c) =\log 3 - S(\rho_b)$ and $H(x|b)=H(x|c)=S(\rho_b)$.

\end{document}